\documentclass[conference]{IEEEtran}
\ifCLASSINFOpdf
\else
\fi

\usepackage{graphicx} 
\graphicspath{ {./figures/} } 

\usepackage{balance}
\usepackage[dvipsnames]{xcolor}         
\newcommand*{\Comb}[2]{{}^{#1}C_{#2}}%

\begin{document}
%
\title{Power to the Data Defenders: Human-Centered Disclosure Risk Calibration of Open Data}

\author{\IEEEauthorblockN{Kaustav Bhattacharjee}
\IEEEauthorblockA{Ph.D. Candidate, Informatics \\ New Jersey Institute of Technology\\
kb526@njit.edu}
\and
\IEEEauthorblockN{Aritra Dasgupta}
\IEEEauthorblockA{Assistant Professor, Data Science \\ New Jersey Institute of Technology\\
aritra.dasgupta@njit.edu}
}


%


\IEEEoverridecommandlockouts
\makeatletter\def\@IEEEpubidpullup{6.5\baselineskip}\makeatother
\IEEEpubid{\parbox{\columnwidth}{
    Symposium on Usable Security and Privacy (USEC) 2023 \\
    27 February 2023, San Diego, CA, USA \\
    ISBN 1-891562-91-6 \\
    https://dx.doi.org/10.14722/usec.2023.237256 \\
    www.ndss-symposium.org, https://www.usablesecurity.net/USEC/
}
\hspace{\columnsep}\makebox[\columnwidth]{}}

\maketitle

\begin{abstract}
The open data ecosystem is susceptible to vulnerabilities due to disclosure risks. Though the datasets are anonymized during release, the prevalence of the release-and-forget model makes the data defenders blind to privacy issues arising after the dataset release. One such issue can be the disclosure risks in the presence of newly released datasets which may compromise the privacy of the data subjects of the anonymous open datasets. In this paper, we first examine some of these pitfalls through the examples we observed during a red teaming exercise and then envision other possible vulnerabilities in this context. We also discuss proactive risk monitoring, including developing a collection of highly susceptible open datasets and a visual analytic workflow that empowers data defenders towards undertaking dynamic risk calibration strategies.

\end{abstract}


%

\section{Introduction}\label{section:introduction}
Open data portals democratize access to hitherto proprietary data. \textit{Data custodians}, like government agencies, can use open data to ensure transparency about their functioning, and \textit{data subjects}, like citizens, can use them to gain insight into the education, healthcare, economic and demographic disparities. However, unrestricted and unchecked access to citizens' data can lead to adverse effects when misused by people with malicious intent. Though these datasets are generally anonymized before release, there are multiple examples where data subjects could be re-identified when these anonymized datasets were linked with other publicly available datasets. 


Researchers showed that 99\% of Americans can be re-identified from heavily anonymized and incomplete datasets using a combination of the demographic attributes~\cite{rocher2019estimating}. In 2016, the Australian Department of Health released \textit{de-identified} medical records for $2.9$ million patients ($10$\% of the population). However, researchers were able to re-identify the patients and their doctors using other open demographic information within a few months~\cite{culnane2017health}. In another example, passengers' private information was disclosed through the public transportation open data released by the city municipal of Riga, Latvia~\cite{lavrenovs2016privacy}. 

These privacy breaches can affect citizens' trust and confidence in the government. People may likely provide false responses to census questionnaires if they think the confidentiality of these responses may be breached~\cite{borgesius2015open}. This calls for a comprehensive study of the possible vulnerabilities present in the open data ecosystem. Multiple studies have discussed disclosures while joining open datasets with private or enterprise ones~\cite{sweeney1997weaving, narayanan2008robust, vokinger2019re, narayanan2006break}. 

In this paper, the scope of our work is confined to the datasets available in the public domain since the open accessibility of these datasets poses a higher risk. As our first contribution, we discuss the curation of high-risk open datasets related to human subjects, along with methods that can detect such vulnerabilities~(Section~\ref{section:methods}). Next, we report these vulnerabilities observed during our ethical hacking exercises into the open data ecosystem~(Section~\ref{section:observations}).
Finding signals of disclosures from a forest of open datasets can be challenging to the defenders of this ecosystem~(data owners and data custodians, henceforth referred to as \textit{data defenders}). Sawyer et al. observed that the performance of human observers deteriorates over time in a low-signal vigilance scenario, which is a likely scenario for data defenders~\cite{sawyer2014cyber}, who are faced with the arduous task of finding needles, i.e., privacy vulnerabilities, in the unsuspecting haystack of linkable open data.
As our third contribution, we discuss how vulnerabilities can be detected and triaged using visual analytic interventions~\cite{bhattacharjee2022privee} that can serve as cognitive aid for data defenders for continuous monitoring of privacy risks.
We focus on the vulnerabilities discovered and their possible remediation through visual analytic solutions~(Section~\ref{section:vis_interventions}). 
We also discuss future work and the challenges that must be addressed to protect open data from disclosure vulnerabilities arising out of highly plausible attack scenarios.
\section{Methods}\label{section:methods}
In this section, we first provide a brief overview of disclosures in the open data ecosystem. This is followed by the methods we used to discover the vulnerabilities and develop a set of datasets that are highly susceptible to disclosures.
\subsection{Background on open data disclosures}
Open datasets can be freely used, re-used, and redistributed by anyone~\cite{opendatadefinition}. The motivation behind creating open datasets is to promote transparency and accountability in public information, especially government data. It helps to democratize information instead of confining it within the data owners and a select few who can pay for it~\cite{kitchin2014towards}. Governments worldwide share these datasets through various open data portals like NYC Open Data~\cite{NYCOpenData}, Chicago Open Data~\cite{CityofChicagoOpenData}, Australian Capital Territory Open Data~\cite{ACTOpenData}, etc., and are generally guided by the FAIR data principle~\cite{wilkinson2016fair}. This principle provides guidelines to improve the \textit{f}indability, \textit{a}ccessibility, \textit{i}nteroperability, and \textit{r}eusability of digital data. 
All these factors make the open data ecosystem a prime choice for research.

However, due to their simple accessibility and findability, these open datasets are generally anonymized before release. Information-theoretic guarantees like k-anonymity~\cite{sweeney2002k}, l-diversity~\cite{machanavajjhala2006diversity}, and t-closeness~\cite{li2007t} are generally applied to these datasets to reduce the possible disclosure risks, i.e., the risk of sensitive information about the individuals mentioned in a dataset being disclosed. Still, joining two anonymized datasets using protected attributes can lead to the disclosure of sensitive information. 
Researchers were able to re-identify $91$\% of all the taxis running in NYC using the NYC taxi open data and a taxi medallion dataset~\cite{douriez2016anonymizing}. The sensitivity of the information contained in this dataset makes it prudent to protect it against all possible disclosures.

But, finding disclosures becomes quite challenging for data defenders since these disclosures can be a function of time. Datasets released at a later point in time may affect the previously released datasets.
Moreover, data defenders follow the practice of ``release-and-forget" where, after a dataset's release, almost no checks are done to ensure the protection of these datasets against newly released datasets~\cite{rocher2019estimating}. Thus, to protect the open data ecosystem from disclosure risks, a collaboration between multiple stakeholders is the need of the hour. Hence, we plan to empower the data defenders to inspect the privacy risks while joining open datasets. 

\subsection{Red team exercise}
In order to explore the vulnerabilities related to the open data ecosystem, we conducted a \textit{red-team exercise} with the help of researchers in data privacy and urban informatics. A red-team exercise can be generally defined as a structured process to better understand the capabilities and vulnerabilities of a system by viewing the problems through the lenses of an adversary~\cite{zenko2015red}. In this subsection, we discuss the different stages of this exercise.
\par \noindent \textbf{Quasi-identifiers and disclosures:} Red-team exercises generally follow the cyber kill chain. It starts with the initial reconnaissance step, where attackers try to find vulnerable entry points into any target system. Moreover, attackers used \textit{quasi-identifiers}~\cite{motwani2007efficient} like age, race, gender, and location to breach privacy by linking multiple datasets~\cite{sweeney2005privacy}. Inspired by this, we bootstrapped our red-teaming activity by searching for datasets with these known quasi-identifiers. During our initial exploration, analysis of these datasets led to interesting observations where some of the datasets have a highly skewed distribution of records across different categories of the quasi-identifiers. Since these datasets have meager number of records for a particular combination of age, race, gender, location, etc., joining them with other datasets can potentially expose sensitive information about these individuals.

\par \noindent \textbf{Disclosures using pairwise joins:} These highly skewed datasets established that vulnerabilities exist in individual record-level datasets. However, this leads to an essential question of whether these datasets can be actually joined with other open datasets to expose sensitive information. Join is a fundamental operation that connects two or more datasets, and joinability is the measure to determine if two datasets are linkable by any number of join keys~\cite{dong2021efficient,chia2019khyperloglog}. When these \textit{join keys coincide with protected attributes} like age, race, location, etc., the outcome of the join can reveal sensitive information about an individual or even disclose the individual's identity. As a next step in the red-teaming exercise, we randomly selected vulnerable pairs of datasets from multiple open data portals~\cite{NYCOpenData, KansasCityOpenData, DallasOpenData} and analyzed them for~\textit{joinability risks} regarding what kind of sensitive information may be leaked.

\textbf{Disclosures using transitive joins:} Inspired by the disclosure examples while joining two datasets and the concept of transitive dependency in databases~\cite{codd1972further}, we explored the concept that two datasets, which have no shared attributes between them, can still be joined if they have shared attributes with a third dataset. Consider that a state's criminal and health records datasets have no common attribute. However, joining them with a particular county records dataset that has shared attributes with both of them can lead to the disclosure of sensitive information. We experimented with different permutations of dataset joins to find an example of transitive disclosure. Though we did not find any examples of transitive disclosure at this stage, this can be an interesting field of research that can further strengthen the inspection of disclosure risk in open datasets. Hence, in another current work, we focus on assessing the risk of disclosure through transition (or \textit{transitive disclosure risk}) in open datasets to prevent the disclosure of sensitive information about an individual or a group of individuals.
\subsection{Data curation exercise}\label{section:dataset_development}

Open data portals contain a multitude of datasets on varying topics like economics, health, and others. However, they may not be relevant in information disclosure about human activity. On top of that, the examples observed during the red teaming exercise press for an urgent need for a smaller subset of open datasets focused on disclosure risks. Hence we curated a seed set of datasets that contains a subset of the open datasets, which may be more susceptible to vulnerabilities related to disclosure. In this subsection, we discuss developing this dataset and the learning outcomes.

\par \noindent \textbf{Data collection:}
Many open data portals are developed using frameworks/APIs like Socrata API~\cite{SocrataAPI}, CKAN API~\cite{CKANAPI}, DKAN API~\cite{DKANAPI}, etc. We selected the Socrata API as our source for the open datasets. Though other APIs could have served a similar purpose, we planned to start with Socrata and develop a generalizable approach that can help integrate the other publicly available APIs. 

First, we queried the list of all available data portals through the Socrata Discovery API. From each of these data portals, we queried the metadata for all the data items available within them. Data items include datasets, maps, data dictionaries, etc. We filtered these results and created a list of $39,507$ datasets.
Manually analyzing all these datasets would be a difficult task for any analyst. However, during our red teaming exercise, we understood that quasi-identifiers' presence could be an indicator of possible disclosures. Hence we developed a semi-automated process that filters datasets if they have some combinations of the known quasi-identifiers like \textit{age}, \textit{sex}, \textit{race}, and \textit{age group}, to name a few. After evaluating the attribute space of the selected datasets, we subsequently updated this list to include more such quasi-identifiers. This helped us to select a broader set of datasets that may be susceptible to disclosure risk through these quasi-identifiers.
\begin{figure}
 \begin{center}
\includegraphics[width=\columnwidth]{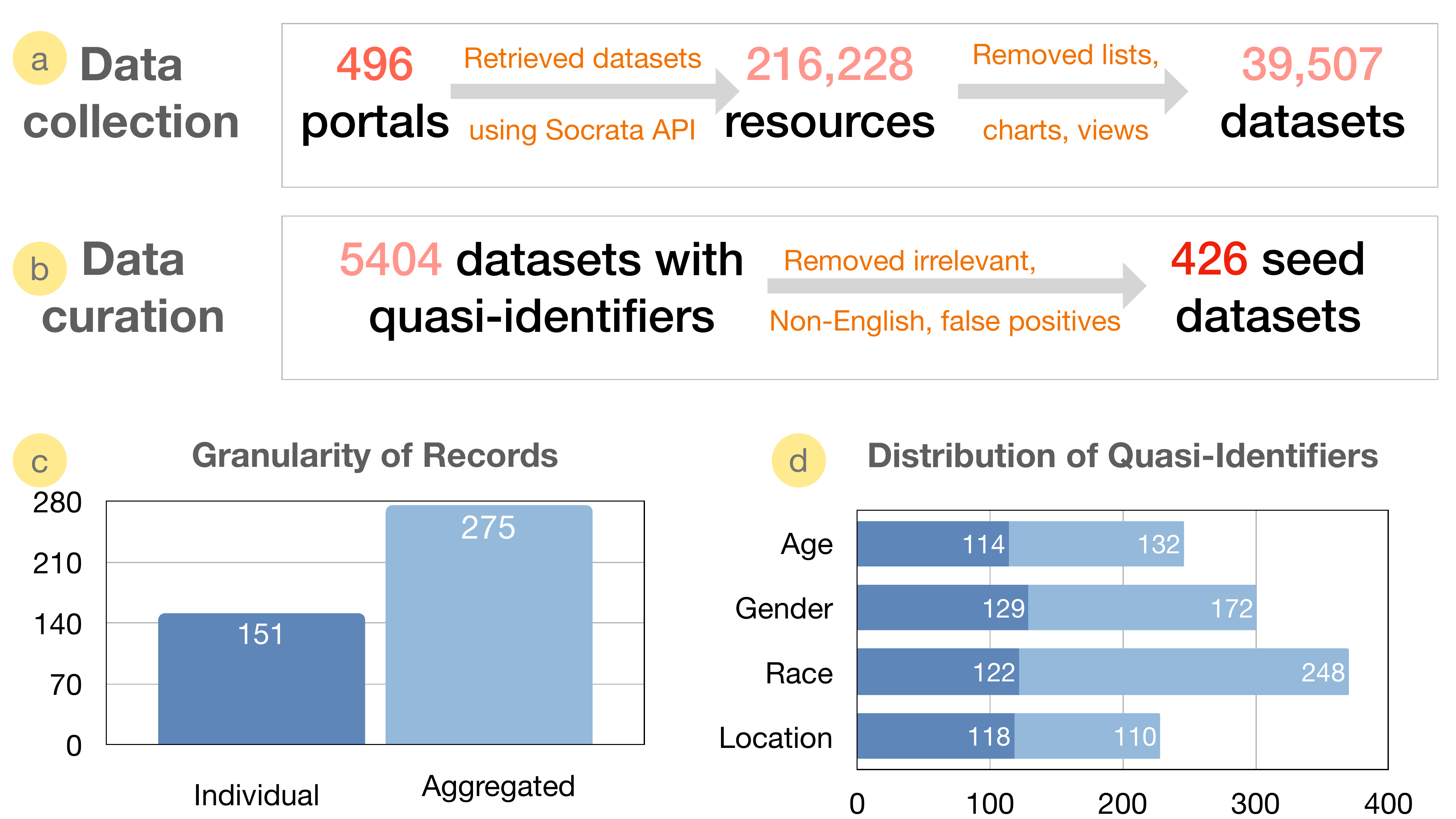}
 \caption{\textbf{Privacy-relevant Data curation:} The dataset development process starts with over $216,000$ data resources from $496$ data portals. After a few filtering steps, it consists of $426$ highly susceptible datasets with different levels of granularities and distribution of quasi-identifiers.}
 \label{figs:dataset_development}
 \end{center}
\vspace{-6mm}
\end{figure}
Multiple iterations of this process led to the development of a set of about $5404$ datasets with some combination of the quasi-identifiers. 
\par \noindent \textbf{Data curation:}
After reducing the set of candidate datasets, the next step was determining if these datasets relate to human subjects and activity.
Hence, we started manually curating the metadata file to understand what each dataset pertains to. For each of the datasets, we opened them in their respective data portals and analyzed them to understand if they were related to human data subjects or not. We observed many such datasets with location attributes (like zip code, address, etc.). However, they do not necessarily relate to human beings, like datasets for \textit{street lamps}, \textit{building details}, etc. We dropped those datasets since they are irrelevant in this context.

Removing these datasets related to non-human objects, we curated a seed set of $426$ datasets of varying granularity. $151$ of these datasets were individual record-level~(e.g., records of people committing crimes) while the rest $275$ datasets were aggregated record-level~(e.g., college records) datasets~(Figure~\ref{figs:dataset_development}). We understand that a dataset collection like this should be continuously updated. In the case of data defenders, they need to be provided with the  infrastructure and techniques to set up data augmentation methods that can fetch and update this collection continuously. 
\begin{figure*}
\vspace{-6mm}
 \begin{center}
\includegraphics[width=\textwidth]{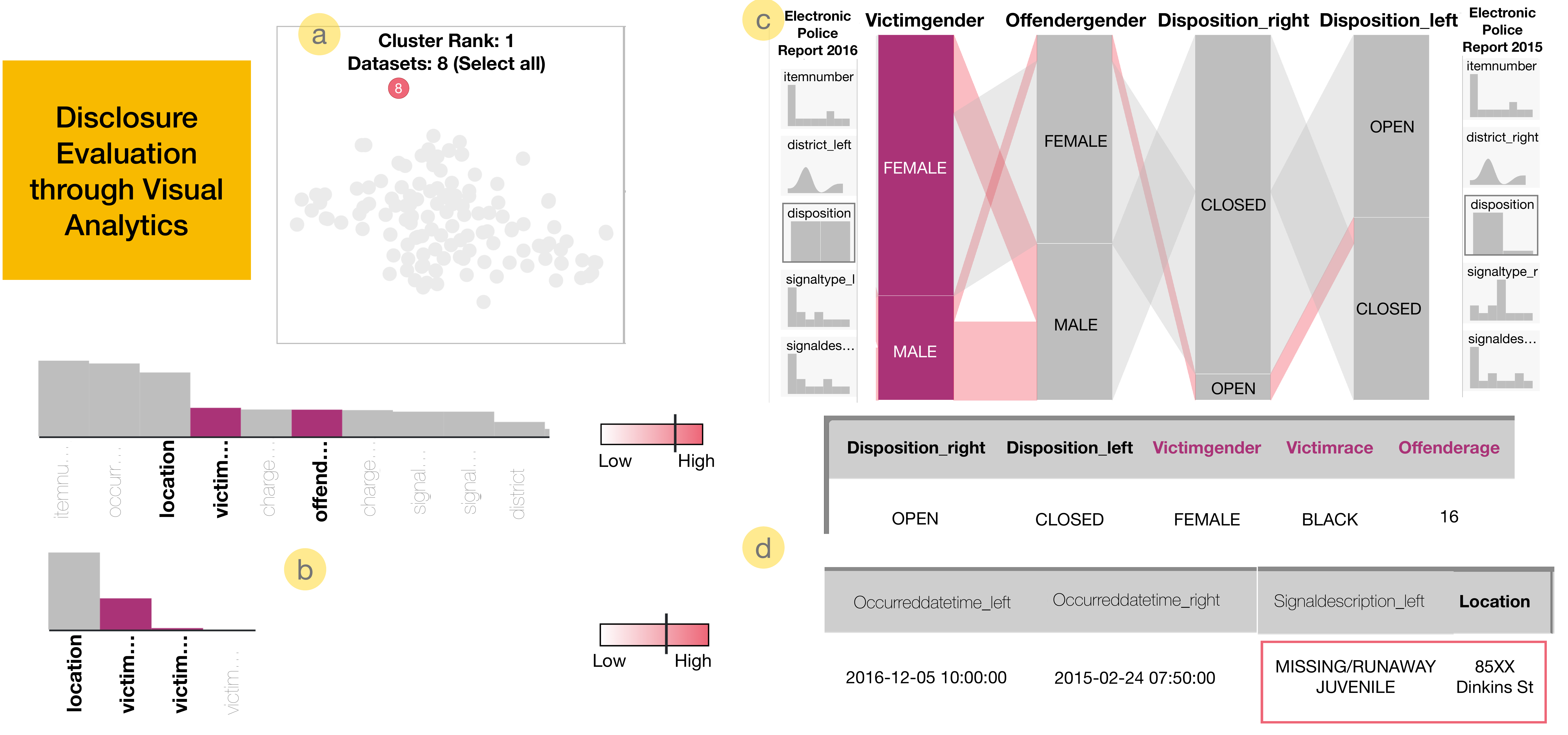}
 \caption{\textbf{Empowering disclosure evaluation through visual analytic techniques: } Using PRIVEE~\cite{bhattacharjee2022privee}, a visual analytic tool for proactive disclosure risk monitoring, data defenders can (a) observe a cluster of joinable datasets formed leveraging their background knowledge, (b) triage the risky dataset pairs based on various combinations of the quasi-identifiers as the join key, and (c) evaluate common records for a particular join key to (d) finally identify disclosures~(Section~\ref{section:vis_interventions}).}
 \label{figs:vis_interventions}
 \end{center}
\vspace{-6mm}
\end{figure*}

\section{Finding Vulnerabilities in Open Data}\label{section:observations}
The red teaming exercise and the set of highly susceptible datasets led to development of a few attack scenarios that the data defenders can emulate to discover vulnerabilities in the open data ecosystem. In this section, we discuss these attack scenarios along with some of the disclosure examples observed. The values reported in this section have been perturbed to a certain extent to protect the data subjects' privacy.
\par \noindent \textbf{Attack exploiting vulnerable entry points:}
Datasets with a highly skewed distribution of records for different categories of a quasi-identifier can serve as vulnerable entry points into the open data ecosystem. For example, the dataset \textit{Whole Person Care Demographics 2}~\cite{WholePersonDemographics2:online} from the \textit{County of San Mateo Datahub} portal~\cite{SMCDatahub:online} had only one record for a 28-year-old female of the Hawaiian race. This can lead to identity disclosure and leak of sensitive information when joined with other datasets. Another dataset, \textit{Demographics for Public Health, Policy, and Planning}~\cite{Demographics:online}, from the same data portal, had only seven records for the age of $18$. However, out of these seven people, only one person was female. This individual can be identified since other identifying attributes like race, language, and city were also present. This may also lead to attribute disclosure if other similar datasets are exploited.
 

Thus, datasets with vulnerable entry points can be exploited to reveal sensitive information about human data subjects. The presence of such datasets in the open data ecosystem is a warning sign that calls for developing a method that acts as the trusted informer for data custodians and informs them of potential disclosures in a proactive manner.

\par \noindent \textbf{Attack using suitable join keys:}
The previous attack scenario established that vulnerabilities exist in individual record-level datasets. These vulnerabilities can be further exploited while joining them with other datasets using suitable join keys. Several iterations of the selection of joinable pairs and join keys led to the discovery of disclosure between the datasets \textit{Juvenile Arrests} and \textit{Adult Arrests} from the \textit{Fort Lauderdale Police Open Data Portal}~\cite{FortLauderdaleOpenData:online}. We observed that two individuals, aged $16$ and $20$, mentioned separately in these datasets, were involved in the same incident of larceny on $10$\textsuperscript{th} March $2018$, at the Coral Ridge Country Club Estate, Fort Lauderdale. This can be an example of identity disclosure by joining two open datasets using a particular join key.
Further investigation revealed other examples where two individuals, aged $18$ and $23$, mentioned separately in these datasets, were involved in the same incident of motor vehicle theft on $18$\textsuperscript{th} of July, $2018$. The presence of linking attributes like \textit{case id} between datasets \textit{Adult Arrests} and \textit{Citations} helped to reveal an incident where a 26-year-old black male, who was arrested for larceny on $27$\textsuperscript{th} September $2021$ at NW $10$\textsuperscript{th} Ave, Fort Lauderdale, was also cited for disobeying stop/yield sign and driving while license being suspended, at NW $8$\textsuperscript{th} Street, just around $3$ miles away from the arrest location. A similar incident was also observed while joining datasets \textit{Citations} and \textit{Juvenile Arrests} on the linking attribute \textit{case id}. In this incident, a 16-year-old white male was first charged with disobeying a red light. He was later arrested for possession of cannabis over $20$ grams on $6$\textsuperscript{th} August, $2015$, both at N Federal Hwy, Fort Lauderdale.

\par \noindent \textbf{Attack exploiting quasi-identifiers:}
We also observed such examples across other open data portals where datasets can be joined using different combinations of quasi-identifiers. Datasets \textit{APD Arrests Dataset by Neighborhood}, and \textit{APD Field Interview Cards Dataset by Neighborhood} from the \textit{Albany Police Department}~\cite{AlbanyPolice:online} were joined on the attributes \textit{age}, \textit{race}, \textit{sex}, and \textit{neighborhoodxy}. We observed that a $24$-year-old white male was interviewed by the police in the Washington Park neighborhood at $08$:$08$ hrs on $2$\textsuperscript{nd} December, $2020$ and was later arrested for trespassing on enclosed property at $11$:$42$ hrs. This leads to an attribute disclosure for the individual arrested as his arrest details are revealed. Joining other datasets like \textit{APD Arrests Dataset by Patrol Zone} and \textit{APD Field Interview Cards Dataset by Neighborhood} from the same data portal revealed a similar incident where a $27$-year-old black female was interviewed at $10$:$22$ hrs on $13$\textsuperscript{th} December, $2020$ and was later arrested at $20$:$27$ hrs for``assault with intent to cause physical injury". In another example, joining datasets \textit{APD Field Interview Cards Dataset by Neighborhood} and \textit{APD Traffic Citations by Neighborhood} on a broader set of quasi-identifiers like \textit{age}, \textit{sex}, \textit{neighborhoodxy} and  \textit{date} led to another interesting observation related to a police incident. We observed that a $22$-year old male was stopped for a field interview on $3$\textsuperscript{rd} January, $2021$ at $1$:$45$ am. Since field interviews are usual routine stop and search activities by the police, this may seem a regular incident. However, the other dataset informed that an individual of the same age and gender received a citation on the same date and at the exact location at $1$:$48$ am, just $3$ minutes after the incident from the first dataset. Since both these records seem to belong to the same person, this is a possible identity disclosure, and it was discovered using a combination of date and quasi-identifiers like location coordinates, age, and gender.

\par \noindent \textbf{Attack leveraging background knowledge:}
Next, we repeated this exercise with added background knowledge about the sensitive attributes used in police datasets and found examples where dataset joins ultimately led to disclosures. For example, two datasets, namely \textit{Electronic Police Report $2016$} and \textit{Electronic Police Report $2015$} from \textit{New Orleans Open Data} portal~\cite{NewOrleans:online}, were joined on quasi-identifiers generally used in police datasets like \textit{victim age}, \textit{offender age}, \textit{victim race}, \textit{victim gender}, \textit{location} and \textit{offender gender}. On inspection of the joined records, we observed that a 23-year-old black male was charged with attempted robbery with a gun against a 29-year-old white male at 6XX Tchoupitoulas St on $12$\textsuperscript{th} July $2015$ at $01$:$00$ hrs and again on $29$\textsuperscript{th} April $2016$ at $03$:$00$ hrs with attempted simple robbery. This is an example of identity disclosure even when \textit{masking techniques} are used on the address. Another observation from these joined records revealed an incident where a runaway female juvenile of age $16$ was reported at 85XX Dinkins St on $24$\textsuperscript{th} February $2015$, and the same incident was closed through a supplemental report one and half years later on $5$\textsuperscript{th} December $2016$. Incidents like these may be rare; hence, identifying the individuals from these records may not be difficult.

\section{Empowering Disclosure Evaluation through Visual Analytic techniques}\label{section:vis_interventions}

The attack scenarios developed using the seed set of datasets highly susceptible to disclosure risks motivated us to explore the visual analytic solution space to understand if the risk can be inspected and communicated to data defenders, leveraging their knowledge through a human-in-the-loop approach. This led to the PRIVEE workflow and interface development, which can guide the data defender toward identifying disclosures using a combination of these attack scenarios. In this section, we discuss how these visual analytic interventions can help the evaluation of disclosures.

During the red teaming exercise,
we randomly selected datasets from various open data portals. However, the candidate datasets can be of the order of hundreds, thus increasing the number of possible combinations. Our collection of highly susceptible datasets has 426 datasets, thus leading to $\Comb{426}{2}$ or $90,525$ possible pairwise combinations. Analyzing all these combinations for disclosure can be a challenging task. Hence, we developed the PRIVEE workflow, leveraging the attack scenarios, which can help the data defender find joinable groups of datasets, triage them based on their risk score and ultimately identify disclosures. Now we discuss these steps using the \textit{New Orleans Open Data}~\cite{NewOrleans:online} portal.

\par \noindent \textbf{Finding joinable datasets leveraging background knowledge:}
The joinability of datasets depends on the presence of shared attributes between the datasets. Hence, developing clusters of datasets based on their attribute space and then understanding the cluster signatures can help find a specific group of highly joinable datasets. 

Suppose data defenders use their background knowledge in criminal history and select some quasi-identifiers popularly observed in police datasets like \textit{victim age}, \textit{victim gender}, \textit{victim race}, \textit{offender age}, etc. In that case, PRIVEE can automatically group the candidate datasets based on their shared attributes~(Figure~\ref{figs:vis_interventions}a). These groups are ranked based on the presence of the selected quasi-identifiers; hence, the first group of datasets would be more relevant based on the user's inputs. PRIVEE also offers insight into the cluster signatures, thus explaining the reason behind the formation of the clusters.

\par \noindent \textbf{Triaging dataset pairs using quasi-identifiers:}
The joinable clusters reduce the number of combinations to be analyzed to a great extent. Selecting the first cluster of $8$ datasets leads to $28$ different pairwise combinations. These datasets can be joined based on a \textit{join key} consisting of some or all of the shared attributes, including the quasi-identifiers. However, during the red-teaming exercise, we realized that analyzing all these dataset pairs based on various join keys can also take considerable time and effort.

PRIVEE attempts to help the data defenders by visualizing all the possible pairwise combinations of the datasets present in a cluster, using a bar chart representing the entropy of the shared attributes. PRIVEE automatically selects some of the shared attributes as the initial join key, giving more preference to the known quasi-identifiers. But the visual cues, like the height of the bars and colored bars representing the privacy-related attributes, help the data defender to make an informed choice~(Figure~\ref{figs:vis_interventions}b). Moreover, these pairs are ranked based on their joinability risk, thus helping the data defenders to focus on highly joinable pairs. PRIVEE also helps to explore the datasets with a highly skewed distribution of records and triage all possible pairwise combinations with these datasets and other individual record-level datasets.

\par \noindent \textbf{Identifying disclosures through suitable join keys:}
The disclosure evaluation process requires the high-risk pairs to be joined using a suitable join key. Multiple iterations of selecting the join key based on the join results can lead to the identification of a disclosure. However, these join results can be hard to interpret regarding the privacy-related attributes and other related attributes.  

PRIVEE presents these results using a modified version of Parallel Sets, a visualization method for the interactive exploration of categorical data that shows the data frequencies instead of the individual data points~\cite{kosara2006parallel}. This helps to understand the relationship between the different attribute categories and identify a specific record with a unique set of attribute values which can lead to disclosures~(Figure~\ref{figs:vis_interventions}c). PRIVEE also offers feature suggestions that can help iterate through the combinations of the shared attributes as the join key. In this case, after examining the feature suggestions, selecting the \textit{disposition} (whether a police incident is open or closed) attribute shows that only one record was open in $2015$ but was closed in $2016$. Further investigation of this record reveals that this is an incident of a runaway female juvenile of age $16$ that was reported at 85XX Dinkins St on $24$\textsuperscript{th} February $2015$, and the same incident was closed through a supplemental report one and half years later on $5$\textsuperscript{th} December $2016$~(Figure \ref{figs:vis_interventions}d). 
Thus, incidents of identity disclosures like this, which were reported during the red teaming exercise, can be identified through the PRIVEE workflow.



\section{Discussion}\label{section:discussion}
Identifying disclosures using traditional search options in open data portals is challenging. Moreover, data custodians might need more information than shown in the search results to find disclosures. Thus, this context demands a visual analytic system specifically targeted toward disclosure evaluation and other privacy pitfalls. PRIVEE can be considered as an initial attempt toward this purpose. The visual analytic design space explored in PRIVEE helps establish a streamlined workflow responsive to the data custodian's inputs yet distilling the results effectively. 

However, this system can have users other than a data custodian. During the development of the PRIVEE workflow, we realized that a data subject could also be interested in discovering if their data can be compromised by exploiting these privacy pitfalls. Our workflow PRIVEE can address the data subjects' perspective too. However, an approach leveraging an individual user's attribute values may be more efficient in this context. Hence, we envision that future design solutions in this space will be more geared toward the data subjects' perspective. This can be incredibly beneficial in encouraging data activism by citizens~\cite{ricker2020open, baack2015datafication, milan2015citizens}.

Another attack scenario we envisaged during the red teaming exercise is the disclosure of sensitive information through the transitive join of open datasets. We are still leading a separate effort toward quantifying the transitive disclosure risk. The primary challenges in this effort are the presence of limited examples yet a high number of possible combinations to explore. This may serve as an important field of research since disclosures like this are difficult to detect by data defenders, yet they can have a massive impact on the privacy of the data subjects. We hope researchers look into different visual analytic solutions to address this attack scenario.  

\section{Conclusion}\label{section:conclusion}
Open datasets are essential in improving government transparency and empowering citizens with access to hitherto proprietary data. We discuss some of the privacy pitfalls of open datasets with real-world examples we observed during an ethical hacking exercise. These examples highlight the importance of addressing these pitfalls on an urgent basis. Towards that end, we develop a collection of highly susceptible datasets and a visual analytic workflow that effectively emulates the strategies developed during the exercise and identifies disclosures. We also envision exploring possible disclosure risks beyond joinable pairs and improving the web-based interface's data processing capabilities in collaboration with big data experts. Since PRIVEE addresses the privacy pitfalls efficiently, this workflow will be used to develop more effective solutions and help data defenders safeguard the interests of the open data ecosystem.
\bibliographystyle{IEEEtranS}
\balance
\bibliography{IEEEabrv,bib}

\end{document}